\documentclass[runningheads]{llncs}

\usepackage[T1]{fontenc}
\usepackage{graphicx}
\usepackage{amsmath,amssymb}
\usepackage{booktabs}
\usepackage{tabularx}
\usepackage{url}
\usepackage[hidelinks]{hyperref}
\usepackage{tikz}
\usepackage{quantikz}

\newcommand{\placeholderfigure}[2]{%
  \IfFileExists{#1}{%
    \includegraphics[width=\textwidth]{#1}%
  }{%
    \fbox{%
      \parbox[c][0.16\textheight][c]{0.92\textwidth}{%
        \centering
        \textbf{Figure placeholder}\\[0.4em]
        #2\\[0.4em]
        \small Expected file: \texttt{\detokenize{#1}}
      }%
    }%
  }%
}

\begin{document}

\title{Variational Quantum Transformer Architecture for Synthetic Language Generation}
\titlerunning{Variational Quantum Transformer Architecture}

\author{
Julian Hager\inst{1}\orcidID{0000-0001-8220-4522} \and
Michael Kölle\inst{1}\orcidID{0000-0002-8472-9944} \and
Gerhard Stenzel\inst{1}\orcidID{0009-0009-0280-4911} \and
Tobias Rohe\inst{1}\orcidID{0009-0003-3283-0586} \and
Jonas Stein\inst{1}\orcidID{0000-0001-5727-9151} \and
Claudia Linnhoff-Popien\inst{1}\orcidID{0000-0001-6284-9286}
}
\authorrunning{J. Hager et al.}

\institute{
Institute of Informatics, LMU Munich, Oettingenstraße 67, 80538 Munich, Germany\\
\email{julian.hager@ifi.lmu.de}
}

\maketitle
\begin{abstract}
We propose a compact NISQ-compatible quantum transformer architecture for synthetic QNLP sequence modelling. The model preserves the autoregressive next-token interface of a classical transformer, but replaces attention and feed-forward sublayers with variational quantum encoder blocks, connector circuits, decoder blocks and a direct two-qubit measurement readout. Token contexts are angle-encoded into small quantum registers, processed by parallel variational heads and encoder integration circuits and conditioned through decoder ancillae to produce a distribution over a four-token vocabulary. We evaluate several architecture variants on deterministic and lexicographic grammar-generation tasks against a compact classical transformer baseline. The quantum models are trainable end-to-end and learn nontrivial grammar structure, including perfect deterministic generation in individual runs and high lexicographic validity in the strongest variant. The classical baseline remains more accurate and stable and the quantum models are sensitive to initialization. The contribution is therefore not a claim of quantum advantage, but a concrete architecture and evaluation of transformer-inspired QNLP sequence modelling under near-term quantum constraints.

\end{abstract}

\section{Introduction}
\label{sec:introduction}
Transformers are the dominant architecture for sequence modelling because they expose a practical autoregressive interface while integrating contextual information through trainable internal blocks~\cite{vaswani2017attention}. Quantum natural language processing (QNLP) offers a complementary perspective in which linguistic structure is represented through tensorial or circuit-like models~\cite{coecke2010mathematical,meichanetzidis2020qnlp}. This raises a natural architectural question: can a transformer-style next-token model be expressed using compact variational quantum circuits while remaining compatible with near-term quantum constraints?\\
This question is nontrivial in the NISQ setting, where useful models must use few qubits, moderate circuit depth and measurement interfaces that can be trained in a hybrid loop~\cite{preskill2018quantum,cerezo2021variational,benedetti2019parameterized}. We address it by proposing a NISQ-compatible quantum transformer for synthetic QNLP sequence modelling. The model maps a fixed context window to a next-token distribution, but replaces classical attention and feed-forward sublayers with variational quantum encoder blocks, connector circuits, decoder blocks and a direct density-matrix readout. The term transformer is therefore used architecturally. The model retains heads, encoder integration, decoder conditioning and autoregressive prediction, but implements these stages with quantum circuits.\\
Related work has also explored quantum natural language generation on near-term devices~\cite{karamlou2022qnlg}. Here we study a complementary architecture-driven setting based on autoregressive variational quantum circuits. We evaluate the architecture on two controlled grammar tasks over a four-token vocabulary: a deterministic cycle, \texttt{AAA BBB CCC}, and a lexicographic language of nondecreasing length-three words. These tasks are deliberately small. Their purpose is to isolate whether the architecture can learn explicit sequence structure under conditions where grammatical validity is measurable. Compared with a compact classical transformer baseline, the quantum variants learn nontrivial grammar regularities and can solve the deterministic task in individual runs, but remain less accurate and less stable overall. Thus this work does not claim quantum advantage. Its contributions are: (i) a concrete variational encoder--connector--decoder architecture for autoregressive QNLP experiments, (ii) parameter-efficient NISQ-scale variants with small active circuit width and (iii) a controlled evaluation using token-level and grammar-level metrics.
\section{Background and Motivation}
\label{sec:background}
\noindent\textbf{Transformers and QNLP.}
A transformer maps token representations to context-dependent features and, in autoregressive use, to next-token probabilities~\cite{vaswani2017attention}. We use this interface as a design pattern rather than reproducing scaled dot-product attention directly. Parallel heads produce intermediate representations, an encoder integrates them, a decoder conditions generation on the encoder state and the final state is measured as a token distribution. This connects to QNLP, where linguistic composition has been studied through tensor-network or circuit-like representations~\cite{coecke2010mathematical,meichanetzidis2020qnlp}, practical near-term executions~\cite{lorenz2021qnlppractice} and software pipelines for mapping language to circuits~\cite{kartsaklis2021lambeq}. Recent work has also explored quantum attention and transformer-like models~\cite{li2024quantum,guo2024quantum,khatri2024quixer}.\\
\noindent\textbf{Variational circuits under NISQ constraints.}
Variational quantum circuits encode data into quantum states, transform them with trainable gates and optimize parameters from measurement-derived outputs~\cite{cerezo2021variational,benedetti2019parameterized}. They are a natural model class for NISQ studies, but they impose architectural pressure by requiring small circuit width and depth together with a readout simple enough for repeated optimization~\cite{preskill2018quantum}. These constraints motivate the choices made below, including small active registers, shallow strongly entangling layers, repeated reduction to two-qubit states and a direct measurement-based output rather than a large learned classical head. The aim is not to replace classical transformers at scale, but to define a compact architecture whose sequence-modelling behaviour can be studied under near-term assumptions.
\section{A NISQ-Compatible Quantum Transformer}
\label{sec:architecture}
The proposed model is an autoregressive hybrid quantum sequence model. Given a context window $x_{t-c:t-1}$, it estimates
\begin{equation}
    p_\Theta(x_t=k\mid x_{t-c:t-1}), \qquad k\in\mathcal{V},
    \label{eq:next-token-distribution}
\end{equation}
where $\Theta$ denotes the trainable circuit parameters. In the experiments, $\mathcal{V}=\{A,B,C,\texttt{space}\}$ and $c=4$. The output vocabulary therefore matches the four computational-basis probabilities of a two-qubit readout state. A detailed circuit diagram is provided in Appendix~\ref{app:circuit-diagram}. This section summarizes the architectural data flow.\\
\subsection{Input Encoding}
\label{subsec:token-interface}
Following the view of quantum models as feature-space methods~\cite{schuld2019quantum}, tokens are mapped to integer identifiers $0,\ldots,|\mathcal{V}|-1$ and angle-encoded into a four-qubit register by applying one rotation per context position,
\begin{equation}
    \lvert\psi_x\rangle
    = \bigotimes_{j=1}^{c} R_X\!\left(\frac{\pi}{2}x_j\right)\lvert 0\rangle,
    \qquad x_j\in\{0,1,2,3\} .
    \label{eq:angle-embedding}
\end{equation}
The wire index represents the token position, so no separate positional encoding is used. For the four-token vocabulary and four-token context, this yields $4^4$ distinguishable context encodings using four data qubits. We write the corresponding density matrix as $\rho_x$.\\
\subsection{Encoder, Connector and Decoder}
\label{subsec:model-overview}
The encoder consists of $E$ blocks. Each block contains two parallel variational head circuits followed by an integration circuit. In the first encoder layer, each head receives $\rho_x$ on the four data qubits and appends two ancilla qubits. Later layers receive the two-qubit state produced by the previous encoder block. Each head applies $V$ strongly entangling layers, implemented with parameterized rotations and CNOT entanglement in PennyLane~\cite{bergholm2018pennylane}, and returns a two-qubit reduced state. The integration circuit combines the two head states with the same variational template and again reduces the result to two qubits. After $E$ blocks, the encoder state $\rho_E$ is therefore a compact two-qubit representation of the complete context.\\
The decoder consists of $D$ blocks and conditions generation on $\rho_E$. In each decoder layer, a connector circuit couples the encoder state to a two-qubit decoder ancilla state. The connector applies CNOT gates from each encoder-output qubit to each decoder-ancilla qubit, followed by one trainable $R_Y$ rotation on each decoder ancilla. The resulting state is combined with the original context state $\rho_x$ and processed by a variational decoder circuit over the two decoder ancilla qubits and four data qubits. The decoder output is again reduced to two qubits and passed to the next decoder layer. This design provides a residual-style path from the original context to every decoder layer while keeping the intermediate representation small.\\
\subsection{Measurement Readout and Variants}
\label{subsec:measurement-output}
After the final decoder layer, the model obtains a two-qubit density matrix $\sigma_D$. Its diagonal entries are used directly as next-token probabilities,
\begin{equation}
    p_\Theta(x_t=k\mid x_{t-c:t-1})
    = \frac{[\operatorname{diag}(\sigma_D)]_k}
           {\sum_{j=0}^{|\mathcal{V}|-1}[\operatorname{diag}(\sigma_D)]_j},
    \qquad k\in\{0,1,2,3\} .
    \label{eq:measurement-probabilities}
\end{equation}
In exact arithmetic the denominator is one. The implementation clamps and renormalizes probabilities for numerical stability. Generation uses greedy autoregressive decoding by appending the most likely token to the context window.\\
All variants use two heads, four data qubits, two-qubit encoder and decoder outputs and a maximum active circuit width of six qubits. Variant names have the form \texttt{q\_e$E$\_d$D$\_v$V$}, indicating encoder depth, decoder depth and variational depth. For the fixed register sizes used here, the number of trainable parameters is
\begin{equation}
    P(E,D,V)=\bigl(48+36(E-1)\bigr)V + D(18V+2),
    \label{eq:parameter-count}
\end{equation}
which counts encoder heads, integration circuits, decoder circuits and connector rotations. The evaluated variants are \texttt{q\_e2\_d2\_v1}, \texttt{q\_e2\_d2\_v2}, \texttt{q\_e2\_d2\_v3} and \texttt{q\_e3\_d3\_v2}. Their parameter counts are reported in Table~\ref{tab:main-results}.
\section{Evaluation on Synthetic QNLP Grammars}
\label{sec:evaluation}
We evaluate the architecture on controlled grammar tasks rather than open-domain language. This keeps the setting small enough for repeated quantum simulation while making structural correctness explicit and measurable.\\
\noindent\textbf{Tasks.}
All experiments use three letter tokens, $A$, $B$ and $C$, plus a separator token, \texttt{space}. Texts are sequences of three-character words separated by \texttt{space} and tokenized at the character level. The deterministic grammar is the periodic word cycle
\begin{equation}
    G_{\mathrm{det}}=(\texttt{AAA},\texttt{BBB},\texttt{CCC})^{*},
\end{equation}
which tests whether the model can learn both word-internal repetition and longer-range phase structure. The lexicographic grammar contains all length-three words with nondecreasing characters,
\begin{equation}
    G_{\mathrm{lex}}=\{abc\in\Sigma^3\mid a\leq b\leq c\} .
\end{equation}
This task admits many valid continuations, so exact token agreement with a sampled test sequence is stricter than grammatical validity.\\
\noindent\textbf{Training protocol and baseline.}
For each grammar, the training text contains 25 words and the test text contains 40 words. Generation starts from the first four test tokens and continues greedily for 100 tokens. We report means and standard deviations over seeds $17$, $23$ and $42$. The same seed controls model initialization and training randomness. For the lexicographic task it also controls the sampled training text, with the held-out test text sampled from the following seed. Quantum variants are trained for 20 epochs with Adam and learning rate $10^{-2}$ using negative log-likelihood loss. The classical baseline is a small encoder--decoder transformer with $d_{\mathrm{model}}=4$, two encoder layers, two decoder layers, two attention heads, feed-forward dimension $2$, dropout $0.1$ resulting in 700 trainable parameters. It is trained for 800 epochs with Adam and learning rate $10^{-3}$. The longer schedule reflects the lower cost of classical batched training and is intended to provide a strong sanity-check baseline rather than a parameter-matched competitor.\\
\noindent\textbf{Metrics.}
We report token accuracy on the generated continuation and a grammar score measuring structural validity. Token accuracy compares generated tokens with the held-out test sequence after the initial context. The grammar score averages four rule-based components like valid-token rate, separator-position rate, word-length validity and either deterministic-cycle validity for $G_{\mathrm{det}}$ or lexicographic-word validity for $G_{\mathrm{lex}}$. The score is a diagnostic complement to token accuracy. In particular, a lexicographic output such as repeated \texttt{AAA} can be formally valid while still having low diversity and low agreement with the sampled target sequence.
\section{Results}
\label{sec:results}
Table~\ref{tab:main-results} reports aggregate performance over the three seeds. Loss values are included for completeness but are most meaningful within a model family because the quantum and classical models use different training schedules. The token and grammar metrics are evaluated under the same autoregressive generation protocol.\\
\begin{table}
    \caption{Performance of the quantum transformer variants and the classical transformer baseline on the synthetic grammar tasks. Values are mean $\pm$ standard deviation over three seeds.}
    \label{tab:main-results}
    \centering
    \scriptsize
    \setlength{\tabcolsep}{3.5pt}
    \begin{tabular}{llrrrr}
        \toprule
        Task & Model & Params & Loss $\downarrow$ & Token acc. $\uparrow$ & Grammar score $\uparrow$ \\
        \midrule
        Det. & Classical & 700 & $0.049\!\pm\!0.020$ & $1.000\!\pm\!0.000$ & $1.000\!\pm\!0.000$ \\
        Det. & \texttt{q\_e2\_d2\_v1} & 124 & $1.041\!\pm\!0.033$ & $0.257\!\pm\!0.045$ & $0.561\!\pm\!0.029$ \\
        Det. & \texttt{q\_e2\_d2\_v2} & 244 & $0.857\!\pm\!0.056$ & $0.567\!\pm\!0.379$ & $0.694\!\pm\!0.271$ \\
        Det. & \texttt{q\_e2\_d2\_v3} & 364 & $0.672\!\pm\!0.085$ & $0.557\!\pm\!0.384$ & $0.755\!\pm\!0.213$ \\
        Det. & \texttt{q\_e3\_d3\_v2} & 354 & $0.926\!\pm\!0.081$ & $0.490\!\pm\!0.207$ & $0.646\!\pm\!0.096$ \\
        \midrule
        Lex. & Classical & 700 & $0.583\!\pm\!0.031$ & $0.557\!\pm\!0.080$ & $1.000\!\pm\!0.000$ \\
        Lex. & \texttt{q\_e2\_d2\_v1} & 124 & $1.298\!\pm\!0.052$ & $0.267\!\pm\!0.029$ & $0.446\!\pm\!0.079$ \\
        Lex. & \texttt{q\_e2\_d2\_v2} & 244 & $1.156\!\pm\!0.048$ & $0.337\!\pm\!0.083$ & $0.532\!\pm\!0.136$ \\
        Lex. & \texttt{q\_e2\_d2\_v3} & 364 & $1.078\!\pm\!0.042$ & $0.280\!\pm\!0.104$ & $0.654\!\pm\!0.188$ \\
        Lex. & \texttt{q\_e3\_d3\_v2} & 354 & $1.154\!\pm\!0.033$ & $0.380\!\pm\!0.156$ & $0.828\!\pm\!0.299$ \\
        \bottomrule
    \end{tabular}
\end{table}\\
\begin{figure}[t]
    \centering
    \includegraphics[width=0.98\textwidth]{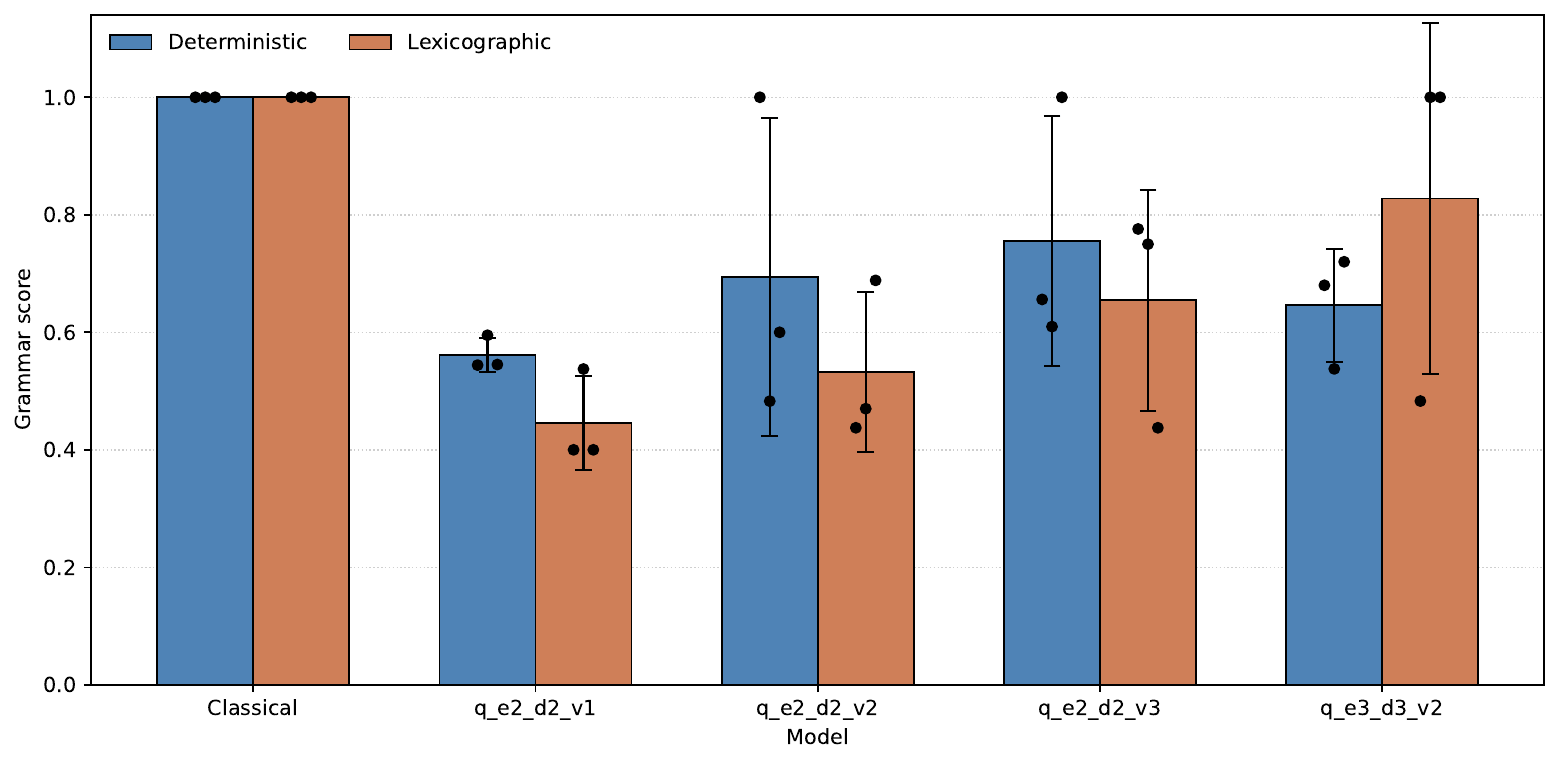}
    \caption{Grammar-level validity of the classical baseline and quantum transformer variants on the deterministic and lexicographic tasks. Bars show mean scores over three seeds, error bars show one standard deviation and black points indicate the individual seed outcomes.}
    \label{fig:grammar-score-bars}
\end{figure}
\noindent\textbf{Deterministic grammar.}
The classical transformer solves the deterministic task in all runs, reaching perfect token accuracy and grammar score. The quantum variants learn weaker but nontrivial structure. The smallest model, \texttt{q\_e2\_d2\_v1}, obtains low token accuracy but a higher grammar score, indicating valid symbols and partial separator alignment without reliable recovery of the full cycle. Increasing variational depth improves performance. \texttt{q\_e2\_d2\_v3} gives the strongest deterministic quantum result with grammar score $0.755\pm0.213$. The large standard deviations are significant. The $V\in\{2,3\}$ variants solve the deterministic continuation perfectly in one seed but fail to do so consistently. Thus the architecture is expressive enough to represent the rule, but the current optimization procedure does not reliably find that solution.\\
\noindent\textbf{Lexicographic grammar.}
The lexicographic task separates exact sequence prediction from grammatical validity. The classical baseline has moderate token accuracy, $0.557\pm0.080$, but perfect grammar score, meaning that it generates valid lexicographic words without reproducing the exact sampled test sequence. Quantum grammar scores increase with circuit expressivity, reaching $0.828\pm0.299$ for \texttt{q\_e3\_d3\_v2}, which also obtains the best quantum token accuracy on this task. However, high lexicographic validity can arise from degenerate outputs such as repeated valid words. The result therefore shows that the architecture can learn formal word constraints, but also that nondeterministic grammars require diversity- or distribution-sensitive evaluation beyond validity alone.\\
\noindent\textbf{Interpretation.}
Across both tasks, the main positive finding is that the quantum transformer is trainable end-to-end and can encode grammar regularities with fewer trainable parameters than the compact classical baseline. The strongest quantum variants use between 354 and 364 parameters, compared with 700 for the baseline. This is not evidence of quantum advantage, since the classical model is clearly more accurate and stable. Rather, the results show that the encoder--connector--decoder circuit design is a viable architecture for controlled autoregressive QNLP experiments, while revealing the optimization instability that future work must address.
\section{Discussion and Limitations}
\label{sec:discussion}
The experiments support this work's central architectural claim, namely that an autoregressive sequence model can be built from variational quantum encoder, connector and decoder blocks while retaining the external next-token interface of a transformer-style model. The positive result is not superiority over a classical transformer, but feasibility. Individual quantum runs solve the deterministic grammar and the lexicographic task shows that the model can learn word-level validity constraints that token accuracy alone does not capture.\\
The comparison between variants suggests that, at the tested scale, increasing variational depth inside each block is more useful than simply adding another encoder--decoder layer. However, the large standard deviations show that training remains sensitive to initialization and optimization. Good solutions appear to exist within the architecture, but the current training procedure does not find them reliably. This is consistent with broader difficulties in variational quantum optimization and should temper any interpretation of the aggregate results.\\
Several limitations remain. The tasks are synthetic and do not test semantic representation, compositional meaning, or natural-language generalization. The experiments use exact simulation rather than shot-based or noisy hardware execution. Finally, the direct two-qubit readout ties the present vocabulary size to four tokens. Larger vocabularies will require more readout qubits, a hybrid output map, hierarchical decoding, or another scalable readout scheme. Future work should therefore study richer encodings, more stable optimization, diversity-sensitive metrics for nondeterministic grammars and execution under realistic NISQ noise and sampling constraints.
\section{Conclusion}
\label{sec:conclusion}
We introduced a NISQ-compatible quantum transformer architecture for synthetic QNLP sequence modelling. The model preserves autoregressive next-token prediction while replacing classical sequence-processing blocks with variational quantum encoder heads, connector circuits, decoder blocks and a direct two-qubit measurement readout.\\
On deterministic and lexicographic grammar tasks, the architecture is trainable end-to-end and learns nontrivial structure, including perfect deterministic generation in individual runs and high lexicographic validity in the strongest variant. The classical baseline remains more accurate and stable, so the results should be read as evidence for a viable architecture, not as a quantum advantage claim. Future work should extend the model to larger vocabularies and contexts, improve optimization stability and evaluate the architecture under shot-based and noisy quantum execution.
\begin{credits}
\subsubsection*{AI Assistance Disclosure.}
The authors used OpenAI ChatGPT for language editing. All experimental design, code, results, interpretation and final responsibility remain with the authors.

\subsubsection{\discintname}
The authors have no competing interests to declare that are relevant to the content of this article.
\end{credits}


\clearpage
\appendix
\renewcommand{\theHsection}{appendix.\arabic{section}}
\section{Detailed Circuit Diagram}
\label{app:circuit-diagram}
The following diagram gives the full circuit-level layout of the architecture instance used to illustrate the model structure. It is moved to the appendix to keep the main paper focused on the architectural definition and experimental results.

\begin{figure}[h]
    \centering
    \includegraphics[width=\textwidth,height=0.72\textheight,keepaspectratio]{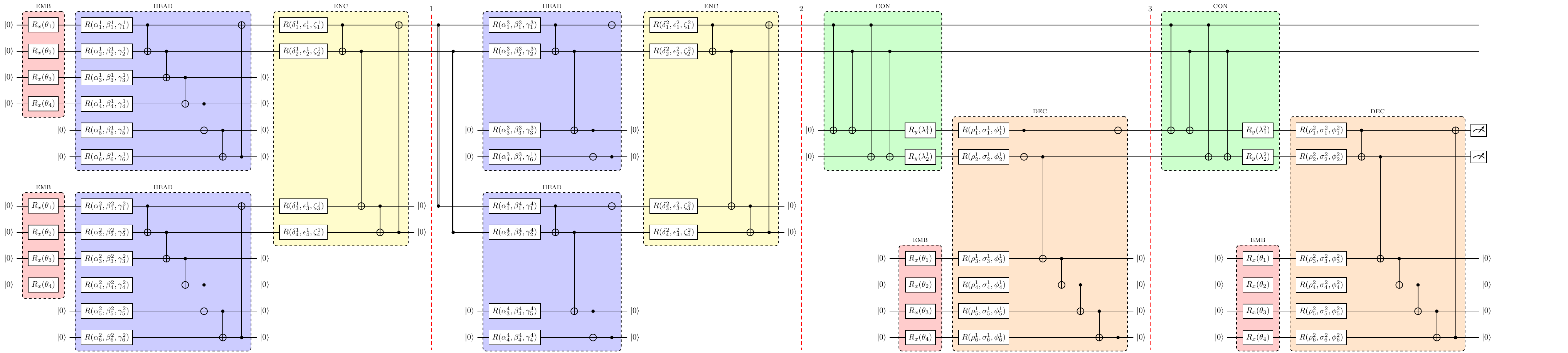}
    \caption{Detailed circuit diagram of the two-encoder, two-decoder quantum transformer instance. Embedding circuits are shown in red, individual heads in blue, encoder integration circuits in yellow, connector blocks in green and decoder blocks in orange.}
    \label{fig:approach_circuit_2Layers}
\end{figure}

\end{document}